# Advancing Interstellar Science:

# A Global Framework for Comprehensive Study of Interstellar Objects

**Authors:**

Omer Eldadi[1], Gershon Tenenbaum[1], and Abraham Loeb[2]

[1]B. Ivcher School of Psychology, Reichman University, Herzliya, Israel

[2]Department of Astronomy, Harvard University, Cambridge, MA, USA

**Abstract**

The operation of the Vera C. Rubin Observatory's Legacy Survey of Space and Time (LSST) marks a transformative moment in humanity's capacity to detect and characterize interstellar objects (ISOs). With projections indicating an increase from a few detections per decade to potentially one every few months, humanity stands at the threshold of unprecedented scientific opportunity offering revolutionary insights into the nature of rocky materials, building blocks of life and technological products from other star systems. This white paper proposes the establishment of the Committee on Interstellar Objects (CIO), a specialized body designed to coordinate global scientific research, maximize observational coverage, and ensure optimal scientific return from these extraordinary objects from outside the solar system through systematic investigation in cosmochemistry, astrobiology, planetary sciences, fundamental physics, advanced technologies and materials science. The proposed framework addresses critical gaps in our current international infrastructure: the absence of coordinated detection, classification and intercept capabilities, insufficient protocols for rapid scientific response and international policy decisions to time-sensitive observations, and the need for effective science communication to maintain government and public support for these ambitious investigations and global threats to Earth. Drawing from successful international collaborations in areas such as the International Space Station (ISS) and the European Organization for Nuclear Research (CERN), CIO would operate through a dual structure: an executive board for time-critical scientific decisions and an expanded committee for comprehensive stakeholder representation. This initiative is not merely aspirational but urgently practical. Recent detections of 1I/'Oumuamua (2017), 2I/Borisov (2019), and 3I/ATLAS (2025) have demonstrated the diversity of ISO characteristics, from 1I/'Oumuamua's unusual acceleration without visible outgassing to Borisov's comet-like behavior, to 3I/ATLAS unusual size and alignment with the ecliptic plane, highlighting the need for comprehensive multi-messenger observations. The window for action is narrow—objects pass through our solar system on hyperbolic trajectories, offering limited observation time before they become permanently inaccessible. Each uninvestigated ISO represents an irretrievable loss of knowledge about stellar nucleosynthesis, protoplanetary disk chemistry, the distribution of organic compounds, and technological relics throughout the galaxy.

# Introduction

Until the past decade, interstellar objects arriving to our vicinity remained in the realm of theoretical constructs, confined to mathematical models and cosmological speculation. This paradigm shifted dramatically with the October 19, 2017 detection of 1I/'Oumuamua, humanity's first confirmed interstellar visitor (Bialy & Loeb, 2018; Meech et al. 2017). The subsequent discoveries of 2I/Borisov in 2019 (Jewitt & Luu, 2025) and 3I/ATLAS in 2025 (Seligman et al., 2025) have transformed interstellar objects from astronomical curiosities to a new frontier of scientific investigation requiring coordinated international collaboration (Hoover et al., 2022; Siraj & Loeb, 2022).

The operational Vera C. Rubin Observatory in Chile, will increase ISO detection capabilities (Dorsey et al., 2025; Hoover, 2022). Conservative estimates suggest the detection of ISO every few months, with some models predicting even higher rates. The dramatic increase in detection frequency fundamentally changes the nature of ISO science from serendipitous discovery to systematic investigation. This transition requires us to consider all possible origins of ISOs, including both natural astrophysical processes and the theoretical possibility of artificial objects from extraterrestrial technological civilizations.

## Scientific opportunities and coordination challenges for ISO research

Interstellar objects present unique scientific opportunities that transcend national boundaries and disciplinary divisions. ISOs traverse our solar system on hyperbolic trajectories, typically remaining observable for only months. Their extreme velocities (so far, up to 60 km/s for 3I/ATLAS outside the solar system) and unpredictable arrival directions make comprehensive study challenging without pre-positioned resources and rapid international coordination (Hibberd et al., 2025).

Technological Requirements: Effective ISO investigation requires integration of data from ground-based telescopes, space-based observatories, and rapid-response scientific missions. No single nation possesses all necessary capabilities, making international cooperation essential rather than optional (Eldadi et al., 2025).

The systematic study of ISOs offers unprecedented scientific returns across multiple disciplines—from planetary science to astrobiology (see Figure 1). Maximizing these opportunities while ensuring equitable access to data and discoveries requires coordinated international frameworks and effective science communication to maintain public support for these ambitious investigations. The discovery of ISOs with anomalous properties, such as 1I/'Oumuamua's non-gravitational acceleration (Micheli et al., 2018) and unusual flat geometry with an extreme aspect ratio (Bialy & Loeb, 2018; Loeb, 2022; Meech et al., 2017), underscores the importance of maintaining scientific openness to all explanatory hypotheses while applying rigorous empirical standards.

**Figure 1**

*Transformative Scientific Opportunities from Interstellar Objects (ISOs)*

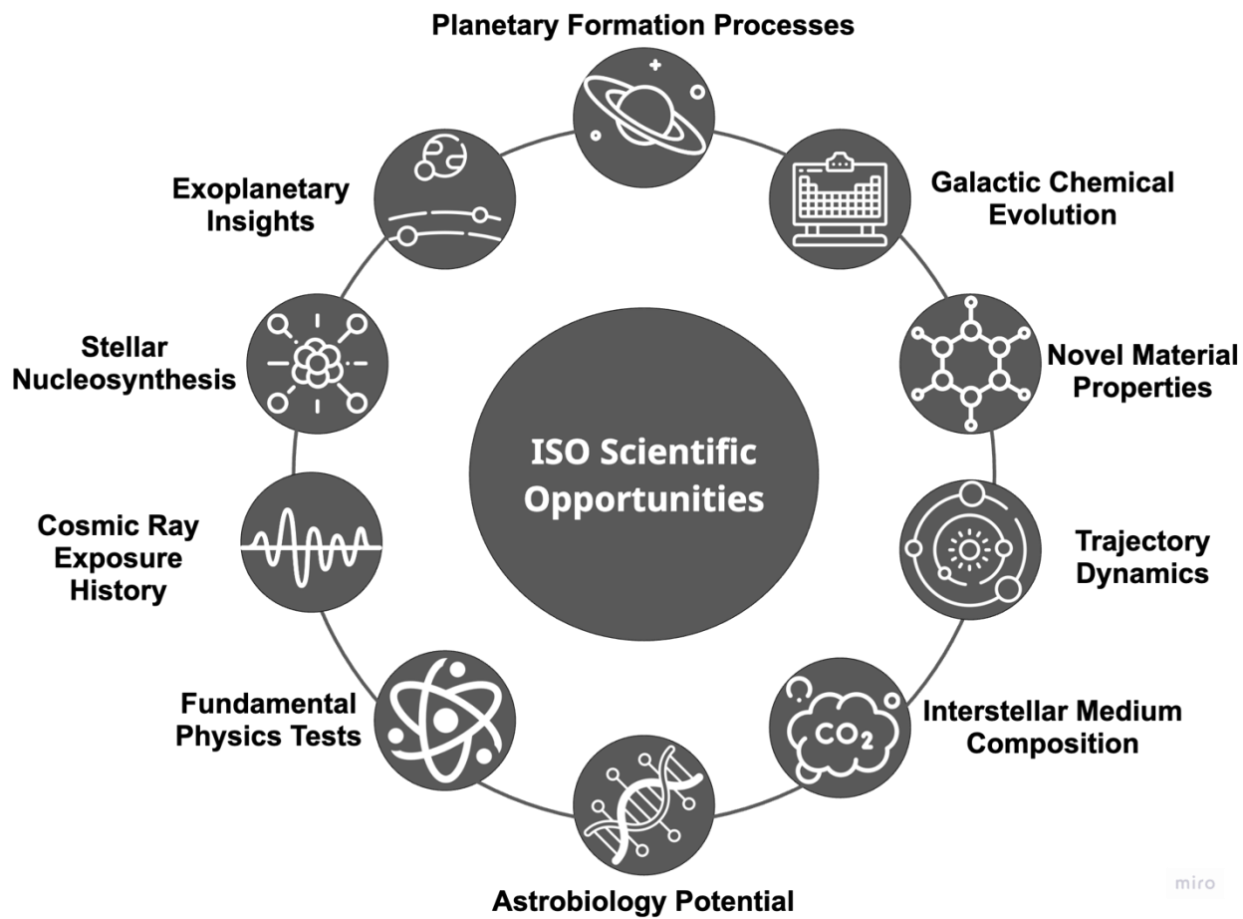

*Note.* The interconnected scientific opportunities presented by systematic ISO investigation. The Vera C. Rubin Observatory's enhanced detection capabilities will enable breakthrough discoveries across multiple scientific domains, from fundamental physics to astrobiology and advanced technologies. Each domain benefits from and contributes to the others, creating multiplicative scientific returns from coordinated international investigation.

**Learning from Precedent**

The international community has successfully achieved remarkable scientific breakthroughs through specialized international frameworks:

(a) The European Organization for Nuclear Research (CERN) demonstrates how nations can pool resources to build and operate scientific infrastructure beyond any single country's capability.

(b) The International Space Station (ISS) partnership shows how rapid decision-making and resource sharing can be achieved in time-critical space operations.

(c) The UN Committee on the Peaceful Uses of Outer Space (COPUOS) provides models for space governance and international cooperation in space science.

(d) The Intergovernmental Panel on Climate Change (IPCC) illustrates how complex scientific assessments can inform policy through coordinated international research.

**Additional successful precedents include:**

(a) The LIGO/Virgo/KAGRA (LVK) and the Event Horizon Telescope (EHT) collaborations united observatories worldwide to achieve impossible-for-one-nation scientific goals.

(b) The Human Genome Project demonstrated how competitive scientific fields can collaborate for transformative discoveries.

(c) The International Ocean Discovery Program shows how nations share costs and benefits of exploring Earth's frontiers. These precedents demonstrate the feasibility of our proposed framework while recognizing the unique aspects of ISO science that require adapted methodologies.

# Current State Assessment

**Scientific Capabilities and Limitations:**

(a) *Detection Infrastructure:* Until 2025 ISO detection relies primarily on wide-field survey telescopes originally designed for other purposes. The Pan-STARRS observatory that detected 1I/'Oumuamua, the Zwicky Transient Facility (ZTF) and the Catalina Sky Survey operate without ISO-specific optimization. The Vera C. Rubin Observatory is dramatically improving detection rates but lacks dedicated follow-up resources;

(b) *Characterization Gaps:* Our ability to determine ISO composition, structure, and origin remains severely limited. Spectroscopic observations of 1I/'Oumuamua were inconclusive (Fitzsimmons et al., 2018), while 3I/ATLAS exhibits unusual spectral properties requiring further investigation. The lack of infrared detection of outgassing or dust from 1/'Oumuamua despite its non-gravitational acceleration remains unexplained (Trilling et al., 2018; Micheli et al., 2018), highlighting our need for more sensitive detection instruments and rapid-response capabilities. Without coordinated multi-wavelength observations and possible sample return missions, fundamental questions about planetary formation and stellar evolution remain unanswered;

(c) *Response Time Constraints:* The median warning time for ISO detection is a few months before closest approach. Current mission planning cycles require years of preparation, making rapid scientific investigation challenging without pre-positioned capabilities.

**Institutional Frameworks**

(a) *Fragmented Coordination:* ISO research currently operates through informal scientific collaborations without binding international agreements or coordinated funding mechanisms.

The International Astronomical Union (IAU) provides nomenclature standards but lacks operational authority for coordinating observations or missions.

(b) *Scientific Communication Gaps:* No established protocols exist for coordinating rapid scientific data sharing and joint publications. The varied interpretations of 1I/'Oumuamua's properties (Bannister et al., 2019) highlight the need for systematic collaborative analysis frameworks.

(c) *Funding Limitations:* ISO research competes with established astronomical priorities for limited resources. The unpredictable nature of ISO arrivals makes sustained funding difficult within traditional grant cycles. Additionally, budget allocations reveal stark disparities - while the Habitable Worlds Observatory (HWO) is slated to receive $11 billion over the next two decades to search for microbial biosignatures (National Academies, 2023), research on interstellar objects that could potentially harbor both biological and technological signatures do not receive substantial federal support.

**Scientific Collaboration and Public Engagement**

Recent research highlights opportunities for enhanced scientific collaboration and public engagement:

(a) *Scientific Collaboration Enhancement:* The rapid international response to 1I/'Oumuamua, despite lack of formal frameworks, demonstrated the scientific community's capacity for self-organization. However, valuable observation time was lost due to coordination delays, highlighting the need for pre-established protocols.

(b) *Public Engagement Opportunities:* Public fascination with ISOs presents unprecedented opportunities for science education and Science, Technology, Engineering, and Mathematics (STEM) engagement. The global interest in 1I/'Oumuamua and 3I/ATLAS generated more public astronomy engagement than any event in the last decade, demonstrating the potential for ISO science to inspire global scientific literacy.

(c) *Modern Communication Channels:* Digital platforms and AI-assisted science communication tools offer new possibilities for real-time data visualization and public participation in ISO science through citizen science programs, creating opportunities for global engagement in scientific discovery.

## Proposed Framework:

## International Astronomical Union Committee on Interstellar Objects (CIO)

### Mandate and Objectives

The International Astronomical Union Committee on Interstellar Objects (CIO) shall serve as the primary international body for coordinating global scientific investigation of interstellar objects. Its mandate encompasses:

*Primary Objectives:*

(i) Coordinate global detection, tracking, and characterization of interstellar objects.

(ii) Develop and maintain rapid research capabilities for high-priority ISOs.

(iii) Establish scientific protocols for ISO investigation and data sharing.

(iv) Coordinate scientific announcements of ISO discoveries.

(v) Maximize scientific returns from ISOs.

(vi) Ensure equitable global access to ISO-derived knowledge.

(vii) Implement the Loeb Scale (Interstellar Object Significance Scale; IOSS) (Eldadi et al., 2025; Trivedi & Loeb, 2025) as the official ISO classification framework, ensuring consistent global risk assessment protocols.

(viii) Develop protocols for potential electromagnetic signaling or physical interaction with ISOs based on their observed characteristics.

*Operational Principles*:

(i) Scientific integrity and evidence-based decision-making.

(ii) Transparency in operations and findings.

(iii) Equitable representation of all nations and stakeholders.

(iv) Rapid research capability balanced with deliberative assessment.

(v) Integration of multiple disciplines including astronomy, planetary science, astrophysics, and astrobiology.

**Organizational Structure**

*Executive Board:* A streamlined decision-making body for sensitive scientific opportunities:

Composition:

- Director-General (proposed: Professor Abraham Loeb, Harvard University).

- Deputy Director for Scientific Operations.

- Deputy Director for International Collaboration.

- Deputy Director for Public Education.

- Representative of the Secretary-General (proposed: Rep. Anna Paulina Luna).

- Rotating Regional Representatives.

- Chief Science Advisor.

*Responsibilities*:

- Determine and validate ISO classification on the Loeb Scale (IOSS) (Eldadi et al., 2025) within 72 hours of initial detection.

- Activate observation protocols for priority ISOs.

- Authorize deployment of rapid-response scientific missions.

- Coordinate real-time observations during ISO passages.

- Coordinate scientific data releases.

- Report to IAU on scientific progress.

*Expanded Committee:*

Composition (35 members):

- Scientific Members (12): Astronomers, astrobiologists, planetary scientists, physicists.

- Psychological and Science Communication Specialists (3).

- Public Education Experts (2)

- Government Representatives (10): Space agencies, research ministries, foreign affairs.

- Civil Society (5): NGOs, educational institutions, public interest groups.

- Private Sector (3): Space industry (SpaceX, Blue Origin, Planet Labs), technology companies.

*Responsibilities*:

- Develop long-term strategic plans.

- Review and approve annual budgets.

- Establish scientific priorities and resource allocation.

- Oversee public education and outreach programs.

- Conduct periodic reviews of CIO operations.

**Technical Infrastructure**

*Global ISO Monitoring System (GIMS):*

(a) Detection Network:

- Integration of existing survey telescopes through data-sharing agreements.

- Dedicated ISO search algorithms optimized for hyperbolic orbits.

- Real-time alert system with <1-hour global notification.

(ii) Characterization Assets:

- Guaranteed time on major observatories (minimum 500 hours annually);

- Dedicated spectroscopic follow-up capabilities.

- Multi-wavelength monitoring capabilities including infrared for outgassing detection.

- Magnetic field sensors for metallic composition assessment.

- Coordinated radar observations for trajectory refinement.

(b) Rapid Scientific Mission Program (RSMP)

(i) Pre-positioned Missions:

- Three standardized research spacecraft maintained in ready status.

- Modular instrument packages for diverse ISO types.

- Launch readiness within 30 days of authorization.

(ii) Mission Profiles:

- Flyby missions for initial characterization.

- Rendezvous missions for extended observation (Hein et al., 2022).

- Sample return missions for highest-priority targets.

(c) Data Management and Analysis Platform:

(i) All ISO observations accessible through unified portal;

(ii) Standardized data formats and metadata;

(iii) Machine learning tools for pattern recognition and classification.

(d) Simultaneous ISO Management Protocol:

Given projected detection rates from the Vera C. Rubin Observatory, CIO must prepare for multiple concurrent ISO passages, a scenario unprecedented in human history, requiring systematic prioritization frameworks.

(i)     Resource Allocation Matrix:

-     Primary Priority Assignment: ISOs classified Level 4+ receive automatic priority status, with resource allocation proportional to classification level.

-     Trajectory-Based Prioritization: Objects with Earth-approaching trajectories (<0.1 AU) receive enhanced monitoring regardless of initial classification.

-     Scientific Value Assessment: For multiple low-level ISOs, prioritize based on: accessibility for intercept missions, unique compositional signatures, potential for sample return, and observational geometry.

(ii)     Operational Deconfliction:

-     Dedicated ISO Coordinators: Assign individual mission directors for each active ISO to prevent resource conflicts.

-     Time-Domain Allocation: Implement rotating observation schedules with automatic handoff protocols between facilities.

-     Rapid Scientific Mission Program (RSMP) Fleet Management: If multiple Level 4+ ISOs present simultaneously, maintain minimum one spacecraft in reserve for highest-priority target.

(iii)     International Burden Sharing:

-     Regional Lead Assignments: Designate primary tracking responsibility based on optimal observation geometry.

- Consortium Activation Protocols: Pre-negotiated agreements for rapid resource pooling when ISO density exceeds single-nation capabilities.

- Emergency Override Authority: Director-General empowered to reallocate committed resources for Level 6+ simultaneous detections.

**ISO Classification Protocol: The Loeb Scale**

We propose the CIO adopts the Loeb Scale as its official classification system. This 0-10 scale provides:

(a) Quantitative thresholds for anomaly and global risk assessment.

(b) Clear escalation triggers for enhanced observation (Level 4+).

(c) Protocols for potential technosignature evaluation.

d) Integration with existing IAU and MPC systems.

(e) Classification triggers automatic response protocols:

(i) Levels 0-1: Routine monitoring.

(ii) Levels 2-3: Enhanced observation allocation.

(iii) Level 4+: Immediate activation of Rapid Scientific Mission Program (RSMP) capabilities.

(iv) Levels 4+: UN Security Council notification.

(v) Levels 8-10: Global emergency protocols.

## Implementation Roadmap

**Phase 1: Foundation (Months 1-6)**

(a) Institutional Establishment:

(i) IAU resolution establishing CIO.

(ii) Appointment of Director-General through IAU procedures.

(iii) Establishment of headquarters at Cambridge, MA, USA, leveraging proximity to the Harvard-Smithsonian Center for Astrophysics and MIT.

(iv) Initial staffing of 25 core positions.

(v) Formal adoption of the Loeb Scale through IAU collaboration.

(b) Preliminary Operations:

(i) Memoranda of Understanding with major space agencies.

(ii) Framework agreements with ground-based observatories.

(iii) Initial budget allocation and financial protocols.

(iv) Communication infrastructure establishment.

(c) Stakeholder Mobilization:

(i) First General Assembly of member states.

(ii) Scientific Advisory Board formation.

(iii) Industry partnership framework development.

(iv) Educational consortium establishment.

**Phase 2: Infrastructure Development (Months 7-18)**

(a) Global ISO Monitoring System (GIMS) Deployment:

(i) Integration of 10 primary observatories.

(ii) Testing of GIMS Alert System with simulated ISO injection into real-time data streams from Pan-STARRS, ZTF, Catalina, ATLAS and Vera C. Rubin, achieving <5% false positive rate.

(iii) Alert system verification exercises.

(iv) Data pipeline optimization.

(b) Rapid Scientific Mission Program (RSMP) Development:

(i) Selection of modular spacecraft platform based on 30-day launch readiness

(ii) Instrument package specifications.

(iii) Launch provider agreements.

(iv) Ground station network establishment.

(c) Capacity Building:

(i) ISO observation training programs.

(ii) Data analysis workshops.

(iii) Science communication certification.

(iv) Citizen science platform development.

(d) Classification Response Protocols Development:

ISO Response Matrix Implementation:

- Levels 0-3: Standard observation queue, routine data collection.

- Level 4-5: UN Security Council notification within 24 hours, Immediate 200+ hours telescope time allocation, activate one RSMP spacecraft to ready status.

- Level 6-7: Full RSMP deployment authorization, activation of Galileo Project assets.

- Level 8+: Implementation of predetermined electromagnetic signaling protocols, G20 emergency session, activation of Article 99 UN Charter provisions.

Protocol Documentation:

- Decision flowcharts for each classification level.

- Legal frameworks for emergency response authorization.

- Communication templates for public announcements.

- International notification chains and timelines.

**Phase 3: Operational Testing (Months 19-24)**

    (a) System Validation:

(i) Full network simulation exercises.

(ii) Asteroid observation as ISO proxies.

(iii) Data flow stress testing.

(iv) International coordination drills.

    (b) Mission Readiness:

(i) First Rapid Scientific Mission Program (RSMP) spacecraft integration.

(ii) Launch rehearsal procedures.

(iii) Sample analysis laboratory certification.

    (c) Performance Benchmarks:

(i) Detection sensitivity:

- Technical specification: magnitude 23.5 for moving objects.

- Practical meaning: Capable of detecting a 50-meter charcoal-dark object at Earth-Sun distance (AU 1).

- Comparison: 2.5x more sensitive than the system that discovered 1I/'Oumuamua.

- Benefit: Provides 3-6 months advance warning for typical ISO encounters.

(ii) Orbit determination: <1000 km uncertainty at 1 AU.

(iii) Alert distribution: <30 minutes globally.

(iv) Data availability: 95% uptime.

**Phase 4: Full Operations (Year 3+)**

    (a) Routine Operations:

(i) 24/7 monitoring coverage.

(ii) Monthly detection exercises.

(iii) Quarterly coordination meetings.

(iv) Annual performance reviews.

(b) Continuous Improvement:

(i) Annual technology review for: quantum sensors for composition analysis, AI-driven anomaly detection upgrading, and laser communication systems for 10x faster data transmission from intercept missions.

(ii) Expanding observatory network.

(iii) Enhanced analytical capabilities.

(iv) Extended mission profiles.

## Scientific Opportunities and Risk Mitigation

**Transformative Science Potential (see Figure 1.)**

(a) Planetary Formation Insights:

(i) Direct sampling of protoplanetary disk remnants.

(ii) Isotopic ratios revealing stellar nursery conditions.

(iii) Size distribution constraining ejection mechanisms.

(iv) Surface composition mapping formation temperatures.

(b) Galactic Evolution Understanding:

(i) Chemical gradients across galactic regions.

(ii) Stellar population mixing timescales.

(iii) Interstellar medium enrichment processes.

(iv) Galactic merger signature preservation.

(c) Astrobiology Implications:

(i) Organic molecule survival in interstellar space (Lingam & Loeb, 2019).

(ii) Panspermia hypothesis constraints.

(iii) Prebiotic chemistry pathways.

(iv) Water and volatile distribution.

**Technical Risk Management**

(a) Detection Completeness:

(i) Challenge: Sky coverage gaps.

(ii) Solution: Partner observatory network expansion.

(iii) Success Metric: 95% sky coverage by Year 3.

(b) Mission Success Rates:

(i) Challenge: Single-opportunity encounters.

(ii) Solution: Redundant spacecraft, robust designs.

(iii) Success Metric: 80% mission success rate.

(c) Data Quality Assurance:

(i) Challenge: Heterogeneous data sources.

(ii) Solution: Standardization protocols, calibration networks.

(iii) Success Metric: 99% data compliance with standards.

**Beyond Technical Risks**

(a) Contamination risks (both directions).

(b) Information security for sensitive discoveries.

(c) Social/psychological impact of high-level detections.

(d) Market/economic disruption scenarios.

(e) Competing national interests in ISO intercepts.

**Operational Excellence**

    (a) International Coordination:

(i) Quarterly virtual coordination meetings.

(ii) Annual in-person conferences.

(iii) Shared operational protocols.

(iv) Cross-training programs.

    (b) Resource Optimization:

(i) Dynamic telescope time allocation.

(ii) Flexible mission architecture.

(iii) Shared technology development.

(iv) Pooled procurement advantages.

    (c) Performance Monitoring:

(i) Real-time dashboard systems.

(ii) Quarterly performance reports.

(iii) Annual independent reviews.

(iv) Continuous improvement protocols.

    (d) Technosignature Assessment:

(i) Implementation of Loeb Scale Level 4-10 criteria.

(ii) Convergent evidence requirements across multiple observables.

(iii) Automatic trigger for enhanced observations at Level 4.

(iv) International notification requirements for Levels 6+.

**Integration with Existing Programs**

The Galileo Project, established at Harvard University in 2021 (Loeb & Laukien., 2023), provides a complementary framework that CIO should formally integrate. The Project's systematic search for extraterrestrial technological signatures through its network of ground-based observatories offers:

*Operational Infrastructure:*

- Existing observatory network with AI-powered anomaly detection.
- Proven protocols for unbiased data collection.
- Established peer review mechanisms.

*Scientific Methodology:*

- Agnostic approach to ISO origin determination.
- Transparent data sharing policies.
- Rigorous classification protocols (Loeb Scale implementation).

*Rapid Response Capabilities:*

- Pre-positioned instrumentation for ISO characterization.
- 24/7 monitoring systems already operational.
- International team of experts on standby.

*Other complementary initiatives include:*

- Breakthrough Listen's radio/optical SETI capabilities.
- ESA's Comet Interceptor mission architecture.
- JAXA's Hayabusa heritage for sample return missions.

CIO should establish formal Memoranda of Understanding with these programs to avoid duplication and maximize scientific return through coordinated observations.

# Global Engagement Strategy

**Scientific Community Integration**

CIO's success depends on creating an inclusive global research ecosystem that maximizes scientific participation while maintaining operational excellence. The Committee will establish comprehensive research opportunities through quarterly open proposal calls, ensuring equitable access to ISO data and observation time. Rather than operating as a closed consortium, CIO will function as a scientific catalyst, amplifying existing research capabilities worldwide.

The collaboration framework will center on dynamic working groups organized around specific ISO encounters, with membership adjusted based on object characteristics and required expertise. When an ISO reaches Level 4 or higher on the Loeb Scale, specialized rapid-response teams will be activated, drawing from pre-qualified researchers across disciplines. Data early access programs will reward contributing institutions while ensuring public release within 6 months, balancing competitive advantage with open science principles.

Knowledge dissemination will occur through annual conferences rotating between continents, with virtual participation ensuring global accessibility. Special emphasis will be placed on real-time data visualization platforms that allow researchers worldwide to monitor ISO observations as they occur, creating a shared global experience of discovery.

**Public Engagement and Science Communication**

The discovery of each ISO represents a unique opportunity to engage humanity in the wonder of cosmic exploration. CIO will implement a tiered communication strategy that scales with the significance of each discovery, from routine detections to potentially paradigm-shifting encounters.

Educational initiatives will span all levels, from K-12 curriculum modules that use ISOs to teach physics and astronomy, to graduate research opportunities at the frontiers of planetary science. Museum and planetarium partnerships will create immersive experiences that allow the public to "ride along" with ISOs through our solar system, using real trajectory data and observations.

Most critically, CIO will maintain a Science Communication Rapid Response Team trained to address public reactions to high-level ISO classifications. When an object reaches Level 4 or above, this team will coordinate global messaging to ensure accurate, consistent information while preventing misinformation spread. Documentary partnerships and social media engagement will maintain public interest between detections, building a sustained constituency for ISO science.

CIO will establish and maintain the Global Interstellar Object Tracker (GIOT) at www.interstellar-objects.org, a comprehensive public portal updated in real-time as new data arrives. This website will serve as humanity's window into ISO science, featuring:

(i) Live Status Dashboard: Current position, velocity, and classification level (Loeb Scale) for all active ISOs, with visual trajectories showing their paths through our solar system.

(ii) Daily Updates Section: Plain-language summaries of new observations, written for general audiences without sacrificing scientific accuracy.

(iii) "What We Know/Don't Know" Format: Transparent presentation distinguishing confirmed data from open questions, preventing misinformation spread.

(iv) Interactive 3D Visualization: Allow users to "fly alongside" ISOs, understanding their trajectories relative to Earth and other planets.

(v) Observation Calendar: When and where ISOs are visible to amateur astronomers, with finder charts for telescope users.

(vi) Classification History: Complete timeline showing how our understanding of each ISO evolved, from initial detection through final classification.

(vii) Alert System: Optional notifications for significant events (new detections, classification changes, perihelion passages).

(viii) Educational Resources: Age-appropriate content explaining ISO science, from elementary explanations to technical papers.

The GIOT platform will update within 6 hours of any significant observation, ensuring the public learns about discoveries simultaneously with the scientific community. During high-interest events like 3I/ATLAS's perihelion passage, the site will provide hourly updates, transforming ISO monitoring from an exclusive scientific activity into a shared human experience. All content will be available in the ten most spoken languages globally, ensuring truly universal access to these cosmic encounters.

**Strategic Partnerships and Resource Coordination**

Effective ISO investigation requires seamless coordination between space agencies, academic institutions, and private sector (SpaceX, Blue Origin and Planet Labs) capabilities. CIO will establish binding Memoranda of Understanding with major space agencies (such as NASA, ESA, CNSA, ISRO) that specify guaranteed response times, resource commitments, and data-sharing protocols.

Academic partnerships will extend beyond traditional astronomy departments to include geology, chemistry, biology, computer, artificial intelligence and engineering programs. The Committee will establish ISO Research Centers at strategically located universities, creating regional hubs for specialized analysis and student training. Private sector engagement will leverage commercial space capabilities, from launch services to advanced sensor development, through performance-based contracts that reward successful ISO characterization.

**ISO Interaction Protocols: Observation, Assessment, and Potential Communication**

Drawing from lessons learned from 3I/ATLAS, whose rare ecliptic alignment, unusual size, a glow in front of the object, detection of nickel but no iron and other unusual characteristics, merit Level 4 classification on the Loeb Scale, CIO establishes a graduated response framework for ISO encounters. This framework recognizes that each ISO represents both a scientific opportunity and a decision point requiring careful assessment before any form of interaction.

*Temporal Decision Framework:*

(i) The Committee adopts a "prudent observation" principle, as exemplified by the 3I/ATLAS approach: intensive monitoring through perihelion before considering any form of electromagnetic signaling. Natural objects typically exhibit enhanced outgassing near the Sun, providing definitive classification. Objects maintaining anomalous characteristics post-perihelion trigger enhanced protocols.

(ii) Key decision points include:

- Initial detection: Classification and resource allocation within 72 hours.
- Pre-perihelion: Maximum observational coverage, no active signaling.
- Perihelion passage: Critical classification window for natural vs. anomalous determination.
- Post-perihelion: Potential communication window if Level 4+ characteristics confirmed.

*Communication Readiness Protocols:*

(i) Should an ISO exhibit characteristics warranting potential communication attempts (Level 4+), CIO will implement a staged approach:

- Stage 1 - Passive Monitoring: Complete electromagnetic spectrum surveillance, pattern analysis for structured signals.

- Stage 2 - International Consultation: 72-hour deliberation period involving UN Security Council, scientific advisory board, and cultural/religious representatives.

- Stage 3 - Message Formulation: If approved, creation of universal mathematical/physical constants-based message, avoiding cultural assumptions.

- Stage 4 - Controlled Transmission: Low-power initial signal from designated facility, with escalation protocols based on response.

*Risk Mitigation Framework:*

Recognizing that ISOs with intelligence might perceive unsolicited communication as threatening, CIO establishes clear boundaries:

(i) No transmission to objects on direct Earth-intercept trajectories.

(ii) Mandatory 30-day observation period before any signaling consideration.

(iii) Automatic communication prohibition for objects showing technological signatures combined with course corrections toward Earth.

(iv) Integration with planetary defense protocols for objects exhibiting hostile characteristics.

*Ethical and Philosophical Considerations:*

(i) The proposed CIO would maintain a standing Ethics Panel including scientists and philosophers to guide decision-making. Planned public notification procedures would

<283-  ensure transparency while preventing panic, with graduated disclosure protocols based on certainty levels.

(ii) The framework explicitly recognizes that, as with 1I/Oumuamua and 3I/ATLAS, we may encounter objects whose nature remains ambiguous despite intensive study. In such cases, CIO defaults to continued observation rather than premature interaction, accepting that some cosmic mysteries may remain unresolved during a single passage. Each ISO encounter, whether definitively natural or tantalizingly ambiguous, contributes to our growing understanding of our cosmic neighborhood and our preparedness for eventual confirmed contact with intelligence from beyond Earth.

## Conclusion

**Seizing the Cosmic Opportunity**

The establishment of CIO represents a natural evolution in humanity's astronomical capabilities, building upon centuries of international scientific collaboration. As the Vera C. Rubin Observatory will transform ISO detections from rare events to routine observations, the international community has a unique opportunity to maximize the scientific return from these celestial objects.

The framework presented here offers a practical and achievable path to coordinated global action. Drawing from successful precedents like CERN, the ISS, and LVK and the EHT, CIO would create the infrastructure necessary to extract maximum scientific value from each ISO encounter. We suggest the following key recommendations:

1. Immediate establishment of CIO under the IAU.

2. Development of a global ISO detection and tracking network.

3. Pre-positioned intercept mission capabilities for rapid deployment.

4. Implementation of a comprehensive ISO classification system to assess potential global threats.

5. Creation of dedicated funding mechanisms through member state contributions.

6. Formal adoption of the Loeb Scale (IOSS) as the standardized 0-10 classification system for all ISOs, providing quantitative thresholds from natural phenomena (Levels 0-3) through potential technosignatures (Levels 4-7) to confirmed artificial origin (Levels 8-10).

7. Integration of science communication and public engagement expertise.

8. Establishment of protocols for electromagnetic and physical interaction with ISOs of potentially artificial origin.

While focused on maximizing scientific return from natural ISOs, CIO's comprehensive design ensures readiness for any unexpected discoveries that may emerge from systematic investigation. The investment required is modest compared to the potential returns. For less than the cost of a single flagship space mission, humanity can establish a permanent capability to study material from other star systems, advancing our understanding of planetary formation, galactic evolution, and the distribution of life's building blocks throughout the cosmos.

The choice is clear: proceed with fragmented national efforts that risk missing critical observations for assessing unprecedented global threats from interstellar space or unite our capabilities to ensure that no ISO passes through our solar system without yielding its scientific information content. The latter path offers not just superior science, but a model for addressing other global challenges through coordinated international action.

# References

Bannister, M., Bhandare, A., Dybczynski, P., Fitzsimmons, A., Guilbert-Lepoutre, A., Jedicke, R., Knight, M., Meech, K. J., McNeill, A., Pfalzner, S., Raymond, S., Snodgrass, C., Trilling, D., & Ye, Q. (2019). The natural history of 'Oumuamua. *Nature Astronomy, 3*, 594-602. https://doi.org/10.1038/s41550-019-0816-x

Bialy, S., & Loeb, A. (2018). Could solar radiation pressure explain 'Oumuamua's peculiar acceleration? *The Astrophysical Journal Letters*, *868*(1), L1. https://doi.org/10.3847/2041-8213/aaeda8

Dorsey, R. C., Hopkins, M. J., Bannister, M. T., Lawler, S. M., Lintott, C., Parker, A. H., & Forbes, J. C. (2025). The visibility of the Ōtautahi-Oxford interstellar object population model in LSST (arXiv:2502.16741). arXiv. https://doi.org/10.48550/arXiv.2502.16741

Eldadi, O., Tenenbaum, G., & Loeb, A. (2025). The Loeb Scale: Astronomical classification of interstellar objects. *arXiv preprint* arXiv:2508.09167. https://doi.org/10.48550/arXiv.2508.09167

Fitzsimmons, A., Snodgrass, C., Rozitis, B. Yang, B., Hyland, M., Seccull, T., ... & Lacerda, P. (2018) Spectroscopy and thermal modelling of the first interstellar object 1I/2017 U1 'Oumuamua. *Nature Astronomy 2*, 133–137. https://doi.org/10.1038/s41550-017-0361-4

Hein, A. M., Eubanks, T. M., Lingam, M., Hibberd, A., Fries, D., Schneider, J., Kervella, P., Kennedy, R., Perakis, N., & Dachwald, B. (2022). Interstellar Now! Missions to explore nearby interstellar objects. *Advances in Space Research, 69*(1), 402-414. https://doi.org/10.1016/j.asr.2021.06.052

Hibberd, A., Crowl, A., & Loeb, A. (2025). *Is the interstellar object 3I/ATLAS alien technology?* arXiv. https://arxiv.org/abs/2507.12213

Hoover, D. J., Seligman, D. Z., & Payne, M. J. (2022). The population of interstellar objects detectable with the LSST and accessible for in situ rendezvous with various mission designs. *The Planetary Science Journal*, *3*(71). https://doi.org/10.3847/PSJ/ac58fe

Jewitt, D. (2025). Interstellar Objects in the Solar System. *Handbook of Exoplanets, 2nd Edition, Springer International Publishing AG, part of Springer Nature.* https://arxiv.org/pdf/2407.06475

Lingam, M., & Loeb, A. (2019). Relative likelihood of success for primitive versus intelligent life. *Astrobiology, 19*, 28. http://doi.org/10.1089/ast.2018.1936

Loeb A. (2022). On the possibility of an artificial origin for 'Oumuamua. *Astrobiology, 22*(12), 1392–1399. https://doi.org/10.1089/ast.2021.0193

Loeb, A., & Laukien, F. H. (2023). Overview of the Galileo Project. *Journal of Astronomical Instrumentation, 12*(01), Article 2340003. https://doi.org/10.1142/S2251171723400032

Meech, K., Weryk, R., Micheli, M., Kleyna, J. T., Hainaut, O. R., Jedicke, R., ... & Keane, J. V. (2017). A brief visit from a red and extremely elongated interstellar asteroid. *Nature*, *552*, 378-381. https://doi.org/10.1038/nature25020

Micheli, M., Farnocchia, D., Meech, K. J., Buie, M. W., Hainaut, O. R., Prialnik, D., ... & Chambers, K. C. (2018). Non-gravitational acceleration in the trajectory of 1I/2017 U1 ('Oumuamua). *Nature 559*, 223–226. https://doi.org/10.1038/s41586-018-0254-4

National Academies of Sciences, Engineering, and Medicine (2023). *Pathways to Discovery in Astronomy and Astrophysics for the 2020s*. Washington, DC: The National Academies Press. https://doi.org/10.17226/26141.

Seligman, D. Z., Micheli, M., Farnocchia, D., Denneau, L., Noonan, J. W., Santana-Ros, T., Conversi, L., Devogèle, M., Faggioli, L., Feinstein, A. D., Fenucci, M., Frincke, T.,

Hainaut, O. R., Hoogendam, W. B., Hsieh, H. H., Kareta, T., Kelley, M. S. P., Lister, T., Marčeta, D., . . . Ye, Q. (2025). Discovery and preliminary characterization of a third interstellar object: 3I/ATLAS. *The Astrophysical Journal Letters, 989.* https://10.3847/2041-8213/adf49a

Siraj, A., & Loeb, A. (2022). The mass budget necessary to explain 'Oumuamua as a nitrogen iceberg. *New Astronomy, 92*, 101730-101734. https://doi.org/10.1016/j.newast.2021.101730

Trilling, D. E., Mommert, M., Hora, J. L., Farnocchia, D., Chodas, P., Giorgini, J., ... & Micheli, M. (2018). Spitzer observations of interstellar object 1I/'Oumuamua. *The Astronomical Journal, 156*(6), 261. https://doi.org/10.3847/1538-3881/aae88f

Trivedi, O., & Loeb, A. (2025). Quantitative mapping of the Loeb Scale. *arXiv preprint* arXiv:2509.06253. https://doi.org/10.48550/arXiv.2509.06253

# APPENDIX A: Proposed Budget Estimation

Initial Investment (Year 1-2):

- Infrastructure Development: $50M

- Rapid Scientific Mission Program (RSMP) Spacecraft (3 units): $300M

- Software/Data Systems: $20M

- Headquarters Establishment: $10M

Total: $380M

Annual Operating Costs:

- Core Staff (25 positions): $15M

- Observatory Time Allocation: $20M

- Mission Maintenance: $30M

- Education/Outreach: $5M

- Administration: $10M

Total: $80M/year

Cost Comparison:

- James Webb Space Telescope: $10B

- CIO 10-year total: $1.08B (11% of a single flagship mission)